# Post-Quantum Key Agreement Protocol based on Non-Square Integer Matrices

HUGO DANIEL SCOLNIK [1, 2, 3, 4] hugo@dc.uba.ar, hscolnik@gmail.com

JUAN PEDRO HECHT [3] phecht@dc.uba.ar, qubit101@gmail.com

*[1]Instituto de Ciencias de la Computación, Universidad de Buenos Aires and CONICET, Buenos Aires, Argentina.*

*[2]Departamento de Computación, Facultad de Ciencias Exactas y Naturales, Universidad de Buenos Aires, Buenos Aires, Argentina.*

*[3]Maestria en Seguridad Informática – Facultades de Ciencias Económicas, Ciencias Exactas y Naturales, Ingeniería, Universidad de Buenos Aires, Buenos Aires, Argentina.*

*[4]Corresponding author*

**Abstract**. We present in this paper an algorithm for exchanging session keys, coupled with a hashing encryption module. We show schemes designed for their potential invulnerability to classical and quantum attacks. In turn, if the parameters included were appropriate, brute-force attacks exceed the (five) security levels used in the NIST competition of new post-quantum standards. The original idea consists of products of rectangular matrices in $\mathbb{Z}_p$ as public values and whose factorization is proved to be an NP-complete problem. We present running times as a function of the explored parameters and their link with operational safety. To our knowledge there are no classical and quantum attacks of polynomial complexity available at hand, remaining only the systematic exploration of the private-key space.

**Contribution of the authors:** both authors contributed equally to this paper.

**Statements and declarations:** the authors declare no competing financial interests. No funding was received for conducting this study.

## 1. Introduction

As it is very well known generating secure key exchange algorithms is a priority for implementing asymmetric protocols [17]. The idea of public key cryptography goes back to the work of James Ellis [8] and the seminal work of Diffie-Hellman [7] which was the first practical solution universally used in SSL, TLS, SSH, IPsec, PKI, Signal, etc.
On the other hand, the imminent appearance of quantum computers able to implement Shor's and Grover's algorithms [2] which seriously affect the currently used cryptographic methods, led to the current research efforts in Post Quantum Cryptography (PQC).
This paper was inspired by E. Stickels's proposals [27] which were cryptanalyzed by V. Shpilrain [21] and C. Mullan [20]. More recently, S. Kanwal and R. Ali [12] published an interesting protocol but it was also cryptanalyzed by J. Liu et al. [15]. A natural alternative was to use rank-deficient matrices but this has been cryptanalyzed by F.Virdia using Jordan canonical forms [29].

It is worthwhile to point out that in the NIST competition for standardization of post-quantum protocols [22], there is none based on the use of non-commutative algebraic systems [21], those dedicated to key exchange protocols (KEP) and their canonical asymmetric cryptosystems, derived using a simple hashing scheme.

This paper aims to provide alternative solutions in this regard. Daniel Brown [5] presented an attack on the early versions of our algorithm [26, v:20221116:142416], which relies on the computation of the characteristic polynomial of the public elements. This approach seems impractical when applied to real-world parameters but led to an updated version using $\mathbb{F}_{2^8}$ field operations [26, v: 20221123:145424]. As this latter kind of attack would be susceptible to a potential Menezes-Wu attack [18] (a fact pointed out by F. Virdia [29]), we reconsider here our first methodology as a usable and secure key-agreement protocol.

## 2. The notation used in this work

p: prime integer, $Z_p$: set of non-negative residuals mod p, products in $Z_p$ (represented by dots), ||: concatenation, Det[A]: the determinant of matrix A, $A^T$: transpose of matrix A, A(i, j): matrix component of the i-th row and j-th column, $\in_{rand}$: random uniform selection in a closed interval, $\oplus$ : bitwise XOR.

## 3. Paper organization

First, we present an overall description of the proposed algorithm and the corresponding protocol, the proof that Alice and Bob will derive a common key, security considerations, and finally some experimental results and a discussion.

## 4. Overall description

The algorithm starts by choosing a prime p shared by Alice and Bob who generate two rectangular matrices each, the first one with more rows than columns and the second one with inverse dimensions, and t is the number of iterations. For each iteration and every entry, a random integer $\in_{rand}$ [(p-1)/2, p -1] is chosen, employing the algorithm given in [6].

Following this scheme, Alice calculates two matrices $A1_k$ and $B1_k$ in each cycle (k=1,…,t) and computes $U_k$ utilizing the matrix product

$$U_k = A1_k \,.\, B1_k \pmod{p} \quad (k=1,2,3,\ldots,t)$$

The vector $U=(U_1, U_2, U_3, \ldots, U_t)$ is sent to Bob. Analogously Bob computes

$$V_k = A2_k \,.\, B2_k \pmod{p} \quad (k=1,2,3,\ldots,t)$$

and the vector $V=(V_1, V_2, V_3,\ldots, V_t)$ is sent to Alice.

We prove that:

$$\text{A-KEY}_k = \text{Det}[A1_k^T \,.\, V_k.\, B1_k^T] \pmod{p}$$

and

$$\text{B-KEY}_k = \text{Det}[A2_k^T \,.U_k.\, B2_k^T] \pmod{p}$$

are equal in each k-cycle.

Finally, Alice computes the hashing of A-CONCAT = $\text{A-KEY}_1$ || $\text{A-KEY}_2$ ||… || $\text{A-KEY}_t$, and Bob the hashing of B-CONCAT = $\text{B-KEY}_1$ || $\text{B-KEY}_2$ ||… || $\text{B-KEY}_t$ which are equal, and hence this is the shared key.

We must observe this protocol is highly parameterizable since we can change the dimensions of the matrices, the number of cycles, the primes, etc. The numerical results (see below) show a very complex shared key can be obtained in a fraction of a second using a standard processor.

## 5. key exchange algorithm

**ALGORITHM 1: PQC multiKEP**

**COMMENTS**
The key Exchange Algorithm (KEP) uses several cycles as defined below.
**INPUT:** see the initial configuration.
**OUTPUT:** shared session key of 512-bits.
**INITIAL CONFIGURATION (PUBLIC VALUES):**
p: a shared prime number that can be obtained randomly.
rows[X|, columns[X]: dimensions of the matrices X:{A, B}, where rowsA=columnsB, columnsA=rowsB and rowsA > columnsA.
rowsA is a value whose maximum is a predefined rowmax value. Our proposal is rowmax=100, rowsA $\in_{rand}$ [5, rowmax] and columnsA $\in_{rand}$ [4, rowsA-1].
t: number of iterations
H( ): hashing SHA3-512.

**ALICE**

**for** k=1 **to** t
    **for** i=1 **to** rowsA
        **for** j=1 **to** columnsA
            $A1_k$ (i,j) $\in_{rand}$ [(p-1)/2, p -1]
        **next** j
    **next** i
    **for** i=1 **to** rowsB
        **for** j=1 **to** columnsB
            $B1_k$ (i,j) $\in_{rand}$ [(p-1)/2, p -1]
        **next** j
    **next** i
    $U_k = A1_k \, . \, B1_k \pmod p$
**next** k
Send the vector $U = (U_1, \ldots, U_t)$ to Bob

```
BOB
for k=1 to t
    for i=1 to rowsA
        for j=1 to columnsA
            A2_k (i,j) ∈_rand [(p-1)/2, p -1]
        next j
    next i
    for i=1 to rowsB
        for j=1 to columnsB
            B2_k (i,j) ∈_rand [(p-1)/2, p -1]
        next j
    next i
    V_k = A2_k . B2_k (mod p)
next k
Send the vector V = (V_1, … ,V_t) to Alice

SESSION KEY OBTAINED BY ALICE

for k=1 to t
    A-KEY_k = Det[A1_k^T . V_k. B1_k^T] (mod p)
next k
A-CONCAT = A-KEY_1 || A-KEY_2 ||… || A-KEY_t
KEY_alice =H( A-CONCAT )

SESSION KEY OBTAINED BY BOB

for k=1 to t
    B-KEY_k = Det[A2_k^T .U_k. B2_k^T] (mod p)
next k
B-CONCAT = B-KEY_1 || B-KEY_2 ||… || B-KEY_t
KEY_bob =H( B-CONCAT )
```

## 6. keys equality proof

**Lemma 1:**
The keys given by Algorithm 1 are equal, that is $KEY_{alice} = KEY_{bob}$

**Proof**: it is very simple taking into account the elementary properties *det(X)=det($X^t$), det(XY)=det(X)det(Y)*, $(XY)^t = Y^t X^t$ where *X, Y* are square matrices of the same dimension We have to prove that for every k (all operations (mod p) )

$Det[A1_k^T . V_k . B1_k^T)= Det[A2_k^T .U_k . B2_k^T]$. Since $V_k = A2_k . B2_k$ and $U_k = A1_k. B1_k$ the keys can be written as follows: $KEY_{alice}= Det[A1_k^T .V_k. B1_k^T) = Det[A1_k^T . A2_k. B2_k . B1_k^T) = Det[(A2_k^T.A1_k)^T. (B1_k.B2_k)^T]$ and $KEY_{bob}= Det[A2_k^T.U_k . B2_k^T]= Det[A2_k^T. A1_k . B1_k . B2_k^T]$ ■

## 7. Derived cipher algorithm

**ALGORITHM 2: PQC multiKEP + simple hashing cipher**

**Observation:** Vector U was received by Bob

Insert here algorithm 1 (up to line 27.)

**BOB CIPHERS A MESSAGE TO ALICE**

1. msg = 512-bit message from Bob
2. **for** k=1 **to** t
3. B-KEY$_k$ = Det[A2$_k^T$ . U$_k$ . B2$_k^T$] (mod p)
4. **next** k
5. B-CONCAT = B-KEY$_1$ || B-KEY$_2$ ||… || B-KEY$_t$

**HASHING CIPHER (C, D):**

6. C = V (* see Algorithm 1 *)
7. D = H[ B-CONCAT ] ⊕ msg
8. Send (C, D) to Alice

**ALICE RECOVERS THE MESSAGE FROM BOB**

9. **for** k=1 **to** t
10. A-KEY$_k$ = Det[A1$_k^T$ . V$_k$ . B1$_k^T$] (mod p)
11. **next** k
12. A-CONCAT = A-KEY$_1$ || A-KEY$_2$ ||… || A-KEY$_t$
13. msg =H[ A-CONCAT ] ⊕ D

The above algorithms can be enhanced by combining modular operations in $Z_p$ with polynomial multiplications in a finite field like $\mathbb{F}_{2^m}$ as in [26], see Discussion below [13, 4]. A convenient choice would be $m$=8 and any of the 30 primitive polynomials of that order [14].

## 8. KEP and HASHING cipher protocols

It is necessary to define a protocol allowing for an interchange of information between Alice and Bob asynchronously to achieve the following objectives:

- deferred communications
- check the integrity of the exchanged information
- mutual authentication to avoid attacks from active adversaries (e.g. man-in-the-middle)
- block replay attacks
- availability of the exchanged information
- perfectly defined formats

The following protocol aims at fulfilling these requirements.

**PROTOCOL: KEP AND CIPHER PUBLIC DATA EXCHANGES**

**INPUT:** any kind of data to be exchanged between entities.
**OUTPUT:** encapsulated message (msg).
**INITIAL CONFIGURATION:**
msg: any kind of information to be exchanged between entities.
Universal-Keyed Message Authentication Code (UMAC): here proposed to assure strong symmetric authentication [3, 11].
ID: any elsewhere predefined and sender-receiver shared identification Tag.
K: sender-receiver shared key.
Tag: a smart label that can store any sort of information from identification numbers as a brief description for each entity. Here the tag is Tag = HMAC-SHA3-512 (HM || Nonce). See Fig 1 and more in [3]
HM : $NH_K(msg_1) \| NH_K(msg_2) \| \cdots \| NH_K(msg_r) \|$ Len, see NH definition in [3].
Nonce: pseudorandom and unique number that changes with each generated tag.
Timestamp: formatted date and time.

**MESSAGE AUTHENTICATION**

1. **Acquire** K and msg
2. **Define** a fixed-length Nonce
3. **Generate** UMAC/SHA3-512/ID/Data
4. **Encapsulate** the concatenation of: [ msg || UMAC/SHA3-512 (HM || Nonce) || Timestamp] into a file
5. **Send** the file and nonce to the receiver

**MESSAGE VALIDATION**

1. **Acquire** at any time the sent file
2. **Recover** msg and UMAC/SHA3-512 (HM || Nonce)
3. **Verify** integrity and sender's identity using K, Nonce, and msg
4. **Accept** or **Dismiss** msg according to the verification result

## 9. Security against brute force attacks

**Lemma 2:**

The complexity of attacking Algorithm 1 by brute force is given by $(ndim\ .\ q)^2.\ t$, where p is the shared prime , $q=\frac{p-1}{2}$ and ndim=rowsA x columnsA.

**Proof:**

To obtain the session key the attacker has to factorize matrices $U_k = A1_k . B1_k$ (mod p), k=1,…,t where the maximum value that can appear in every entry (i, j) is p-1, independently of ndim (see Algorithm 1). Thus, defining $q=\frac{p-1}{2}$; for each entry of each matrix, it is necessary to try q values, that is $(ndim.q)^2$ possibilities. The attack is successful only if the factorization can be obtained for each k=1,…,t, something that can be easily avoided by choosing appropriate parameters ■

Example:
If p = 2147483647, ndim=100 x 90, t =10, then the complexity is ~ $2^{74}$

## 10. Security against analytical attacks

It must be stated that $U_k$, $V_k$ are always singular matrices, therefore blocking inversions and subsequent linear attacks. This is a consequence of the fact that the rank of the product of two (rectangular) matrices must be less or equal to the minimum rank of each one, therefore the rank of the product $U_k$ (or $V_k$) of dimension rows x rows (rows > cols) is less than rows, so its determinant is zero.

The core of the function defining the session keys (lines 30 and 35 of Algorithm 1) comprises the product of four rectangular matrices, a public component ($U_k$,$V_k$) flanked by two private matrices ($AN_k$, $BN_k$, N={1,2}).
The determinant is distributive concerning this product, and rearranging the product to be the multiplication of two public elements ($U_k$ . $V_k$). As mentioned before, it does not work over real-life parameters. This fact forces the factorization of the ($U_k$,$V_k$) components, which is computationally hard.

The factorizations $U_k = A1_k . B1_k$ (mod p), k=1,…,t can be solved in $U_k \in R^{n\times n}$, $A1_k \in R^{n\times m}$, $B1_k \in R^{m\times n}$ (real realm), using the SVD (Singular Value Decomposition) if there are no restrictions regarding the nonnegativity of the factors, but if they are imposed as in our proposal, then Vavasis and references therein [28] proved that the problem is NP-hard in the continuous case. For the Boolean case, N. Gillis [10] wrote: *"…for exact factorizations, the rank-one problem is trivial. For higher ranks, Boolean factorization is equivalent to finding a rectangle covering of the matrix U. This is equivalent to the so-called biclique problem (given a bipartite graph defined by U, find the smallest number of complete bipartite subgraphs that cover the graph) which is NP-complete as proved in* [24]. *(sic)".*

The Boolean NP-complete complexity was formally proven by Miettinen and Neumann [19].

## 11. Semantic security of algorithm 2

Here we provide an informal view of this aspect, based on concepts mostly derived from Bellare's work [1]. In a nutshell, semantical security measures the resistance of any encryption algorithm to attacks using chosen plaintext or ciphertext selected by the attacker, who has access to the encryption and decryption modules working as oracles; without knowledge of the key selected for enciphering [1]. The semantic security term is strongly related to other definitions: the one-way functions and the non-malleability of ciphertexts.
Indistinguishability under chosen plaintext attack (represented as IND-CPA) is equivalent to the property of semantic security and is considered a basic requirement for most provably secure public-key cryptosystems [1]. One-way refers to bidirectional functions that have a probabilistic-polynomial time algorithm that converts domains into codomains, but no such algorithm is known that inverts the procedure. Non-malleability refers to the resistance to modify slightly the ciphertext to obtain meaningful recovered plaintext [1]. The next concept to define is the indistinguishability of different ciphertexts of two similar but different plaintexts, an attacker could not assign a ciphertext of one of them to any one of the plaintexts. This feature is generally presented as a game between a challenger (the algorithm defender) and an adversary (the algorithm attacker) [1]. The challenger generates a key pair PK, SK (public key and secret key respectively), based on any security parameter k (which can be the key size in bits), and publishes PK to the adversary. The challenger retains SK. Here we describe the adaptative version of the game. The adversary may perform any number of encryptions, decryptions, or any other operations. (The adversary is a probabilistic polynomial Turing Machine) [1]. Eventually, the adversary submits two distinct chosen plaintexts $m_0$ and $m_1$ to the challenger (of the same length). The challenger selects a bit $b \in \{0, 1\}$ uniformly at random, and sends the challenge ciphertext $C = E(PK, m_b)$ back to the adversary. The adversary is free to perform any number of additional computations or decryptions (except C, this step is the adaptative phase of the attack). Finally, its outputs in polynomial time a guess for the value of b [1]. The adversary wins the game if it guesses the bit b, and winning means the algorithm is not indistinguishable and secure, else the algorithm reaches the strongest available security level: IND-CCA2. (Indistinguishable chosen ciphertext adaptative attack). Formally, a cryptosystem is indistinguishable under an adaptative chosen ciphertext attack if no adversary can win the above game with probability p greater than $1/2 + \epsilon_k$, where $\epsilon_k \leq 1/\Pi^K$ ($\Pi^K$ arbitrary polynomial function) and $\epsilon_k$ is defined as a negligible function in the security parameter k [1]. For Algorithm 2 we prove the IND-CPA security level and explain how it could be easily adapted to reach the IND-CCA2 security level. The use of the UMAC function [3] in our Protocol fills this need in such a way that the practical implementation of Algorithms 1 and 2 culminates with the desired provable-security level, accepting the random-oracle characteristic [16] provided by UMAC.

## 12. NIST PQC security level of algorithm 2

NIST bases its classification on the range of security strengths offered by the existing NIST standards in symmetric cryptography, which NIST expects to offer significant resistance to quantum cryptanalysis. In particular, NIST defines a separate category for each of the following security requirements [23]. As previously described in the brute-force attack section, reasonable parameters exceed largely Level-1 PQC security.

## 13. A toy numerical example

**Shared parameters**: prime p = 5303, rowsA=3, columnsA=2, t=2

## ALICE

| **Iteration 1** | **Iteration 2** |
|---|---|
| Alice $A1_1$ = | Alice $A1_2$ = |
| 1123 341<br>14 238<br>1041 13 | 665 1338<br>622 38<br>505 1617 |
| Alice $B1_1$= | Alice $B1_2$= |
| 1525 1019 1561<br>1561 716 862 | 925 1412 598<br>364 463 409 |
| Alice $U_1$ = | Alice $U_2$ = |
| 1707 4410 5290<br>446 4372 4284<br>1009 4184 2883 | 4436 4695 978<br>549 4954 379<br>416 3406 3500 |

## BOB

| **Iteration 1** | **Iteration 2** |
|---|---|
| Bob $A2_1$ = | Bob $A2_2$ = |
| 802 2435<br>1206 3408<br>707 3723 | 656 13<br>1900 107<br>611 1537 |
| Bob $B2_1$= | Bob $B2_2$ = |
| 1174 2805 1242<br>3110 814 550 | 2192 1270 845<br>820 1022 2194 |
| Bob $V_1$ = | Bob $V_2$ = |
| 3083 5209 2014<br>3429 159 4847<br>4831 2322 3791 | 893 3229 4815<br>4837 3429 117<br>1182 2858 1374 |

Alice computes A-KEY$_k$ = Det[$A1_k^T . V_k . B1_k^T$] (mod p) for k=1, 2

A-KEY$_1$ = 3207
A-KEY$_2$ = 2121
**Alice's key** =
*0c3322f92446b51e3372d2a7bd2b81265bb96f32fa38562e4c02414e3c73d85ca4b358363b8792461d4033c1d76 23589c0f6c07ab01e33b6a7294019e125c779*

Bob computes B-KEY$_k$ = Det[$A2_k^T . U_k . B2_k^T$] (mod p) for k=1, 2

B-KEY$_1$ = 3207
B-KEY$_2$ = 2121
**Bob's key** =
*0c3322f92446b51e3372d2a7bd2b81265bb96f32fa38562e4c02414e3c73d85ca4b358363b8792461d4033c1d76 23589c0f6c07ab01e33b6a7294019e125c779*

## Example of the cipher algorithm

**Bob ciphers a message to Alice**
B-KEY = 32072121
PLAINTEXT msg (formatted as a 64-byte string with spaces appended on the right) = "***This is a secret communication.***"
CIPHERTEXT C =

| | | | | | |
|---|---|---|---|---|---|
| 3083 | 5209 | 2014 | 893 | 3229 | 4815 |
| 3429 | 159 | 4847 | 4837 | 3429 | 117 |
| 4831 | 2322 | 3791 | 1182 | 2858 | 1374 |

CIPHERTEXT D =
(88,91,75,138,4,47,198,62,82,82,161,194,222,89,228,82,123,218,0,95,151,77,56,71,47,99,53,39,83,29,246,124,132,147,120,22,27,167,178,102,61,96,19,225,247,66,21,169,224,214,224,90,144,62,19,150,135,9,96,57,193,5,231,89)

**Alice recovers the message**
A-KEY = 32072121
RECOVERED msg = “***This is a secret communication.***”

## 14. Numerical experiments

Programmed in RUST
OS: Windows 10 Pro (64 bits)
Processor: Intel(R) Core(TM) i7-3770K CPU @ 3.50GHz
RAM: 8,00 GB
The CPU times reported in the following Table 1 are the mean values of 10 runs for each combination of the variables, prime p = $2^{31}$-1= 2147483647 (~31 bits)

| rowsA | columnsA | Cycles | Complexity | CPU time in milliseconds |
|---|---|---|---|---|
| 5 | 4 | 10 | $\sim 2^{72}$ | 0.94 |
| 5 | 4 | 20 | $\sim 2^{73}$ | 1.15 |
| 5 | 4 | 100 | $\sim 2^{75}$ | 2.69 |
| 6 | 5 | 10 | $\sim 2^{73}$ | 0.99 |
| 6 | 5 | 20 | $\sim 2^{74}$ | 1.39 |
| 6 | 5 | 100 | $\sim 2^{76}$ | 3.95 |
| 20 | 19 | 10 | $\sim 2^{80}$ | 8.92 |
| 20 | 19 | 20 | $\sim 2^{81}$ | 13.45 |
| 20 | 19 | 100 | $\sim 2^{84}$ | 74.47 |
| 100 | 99 | 10 | $\sim 2^{89}$ | 755.38 |
| 100 | 99 | 20 | $\sim 2^{91}$ | 1503.54 |
| 100 | 99 | 100 | $\sim 2^{93}$ | 7488.44 |

**Table1.** Algorithm 1 complexity and throughput time as a function of the parameters.

Using 128 bits integers in RUST with prime p = 18446744073709551113 (~64 bits):

| rowsA | columnsA | Cycles | Complexity | CPU time in milliseconds |
|---|---|---|---|---|
| 5 | 4 | 10 | $\sim 2^{138}$ | 1.29 |
| 6 | 5 | 10 | $\sim 2^{139}$ | 0.98 |
| 20 | 19 | 10 | $\sim 2^{146}$ | 26.84 |
| 100 | 99 | 10 | $\sim 2^{156}$ | 3203.48 |

**Table 2.** Algorithm 1 complexity and throughput time as a function of the parameters.

## 15. Discussion

Algorithm 1 has been implemented in different computer languages and shows that extremely high complexity can be achieved in fractions of a second on a standard I7 processor. The fact that by modifying the input variables (number of rows, columns, primes, iterations) practically any security level can be easily obtained without resorting to multiple precision, leads to very fast implementations. Depending upon the computer architecture and software implementation, larger primes can be used for reaching higher complexity levels.

An enhancement is to use $\mathbb{F}_{2^m}$ polynomial fields [26], recurring to very fast field multiplications based on hard-coded discrete logarithm tables of any field generator base, on some steps. In such a way the lack of uniformity in the operations constitutes an additional barrier to possible algebraic attacks. This proposal could be very easily implemented without downgrading performance [4].

It is particularly important to use the UMAC function in the Protocol because it is similar to Merkle's trees for PQC digital signatures [2] and also plays the role of achieving maximal semantic security [1] and simultaneously strengthens its post-quantum character. **Note:** as Black et al. [3] state "*the security of UMAC is rigorously proven, in the sense of giving exact and quantitatively strong results which demonstrate an inability to forge UMAC-authenticated messages assuming an inability to break the underlying cryptographic primitive. (sic)*"

## 16. Conclusions

The algorithm presented in this paper is such that very high complexity can be reached using small primes, normal precision, and small rectangular matrices, leading to very fast computer implementations.

**Acknowledgments**: to D. R. L. Brown, D. B. Szyld, L. Liberti, N. Gillis, F. Virdia, J. Di Mauro, I. Córdoba, S. Barzola, G. Cucatti and Y. Alis for many interesting discussions, theoretical insights and computer implementations.